\documentclass{article}

\newcommand{\documentdate}{\today}  

\usepackage{mystyle,varioref,subfigure,graphicx}
\usepackage{a4wide,color}

\usepackage{latexsym,amsmath,amstext} 
\usepackage[english]{babel}          
\usepackage[utf8]{inputenc}     
\usepackage{multirow}
\usepackage{multicol}
\usepackage{tabularx}

\usepackage{natbib}

\usepackage{endnotes}

\usepackage{authblk}

\newcommand{\mc}[2]{\multicolumn{#1}{|>{\setlength{\hsize}{#1\hsize}}X|}{#2}}
\newcommand{\argmax}{\mathop{\mathrm{arg\,max}}}

\usepackage{chngcntr}
\counterwithout{figure}{section}
\counterwithout{table}{section}
\counterwithout{equation}{section}

\let\footnote=\endnote

\date{\documentdate}

\title{Interaction prediction between groundwater and quarry extension using  discrete choice models and artificial neural networks}

\author[1]{Johan Barthélemy\thanks{johan\_barthelemy@uow.edu.au, corresponding author}}
\author[2]{Timoteo Carletti}
\author[3]{Louise Collier}
\author[3]{Vincent Hallet}
\author[2]{Marie Moriamé}
\author[2]{Annick Sartenaer}

\affil[1]{SMART Infrastructure Facility, University of Wollongong, NSW, Australia}
\affil[2]{Namur Research Center for Complex Systems, University of Namur, Belgium}
\affil[3]{Department of Geology, University of Namur, Belgium}

\begin{document}

\maketitle

\begin{center}
The final publication is available at Springer via http://dx.doi.org/10.1007/s12665-016-6268-z
\end{center}

\begin{abstract}
Groundwater and rock are intensively exploited in the world. When a quarry is deepened the water table of the exploited geological formation might be reached. A dewatering system is therefore installed so that the quarry activities can continue, possibly impacting the nearby water catchments. In order to recommend an adequate feasibility study before deepening a quarry, we propose two interaction indices between extractive activity and groundwater resources based on hazard and vulnerability parameters used in the assessment of natural hazards. The levels of each index (low, medium, high, very high) correspond to the potential impact of the quarry on the regional hydrogeology.

The first index is based on a discrete choice modelling methodology while the second is relying on an artificial neural network. It is shown that these two complementary approaches (the former being probabilistic while the latter fully deterministic) are able to predict accurately the level of interaction. Their use is finally illustrated by their application on the Boverie quarry and the Tridaine gallery located in Belgium. The indices determine the current interaction level as well as the one resulting from future quarry extensions. The results highlight the very high interaction level of the quarry with the gallery.
\end{abstract}





\section{Introduction and motivation}

\begin{figure}[ht]
  \centering
  \includegraphics[scale=0.24]{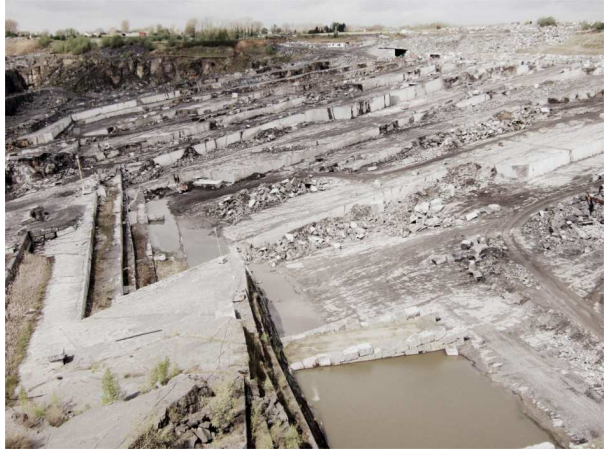}
  \caption{The Neufvilles (Clypot) quarry site (Walloon Region, Belgium). The quarry has a maximal dewatering flow of 2.000.000$m^3$ per year.  The water table dropped from an initial altitude of 70$m$ down to currently 25$m$. Picture courtesy of Louise Collier.}
  \label{fig_quarry_neufvilles}
\end{figure}

There are two underground resources intensively exploited in the world: groundwater and rock. Given the population density and environmental pressures, quarry lateral extension may be limited. Hence, the only solution for the rock operators is to excavate deeper as long as the deposit structure makes it possible. In this context, the aquifer level of the exploited formation is often reached and pumping systems have to be installed to depress the water table below the quarry pit bottom. For instance there are about 60 quarries out of 160 being currently active in Wallonia (the southern region of Belgium) performing dewatering such as the Neufvilles (Clypot) exploitation site illustrated in Figure \ref{fig_quarry_neufvilles} (\citealt{CollHall13}).  As a result, this affects the regional hydrogeology and, in some cases, the productivity of the water catchments could be threatened.

Generally, the evaluation of environmental impacts requires for public and private projects to identify, describe and assess the direct and indirect effects, in the short, medium and long term, of the implementation and the establishment of the project. This assessment takes into account components such as human, fauna, flora, soil, water, air, climate and landscape as well as economical issues (\citealt{Spw14}). Then, in order to assess the interaction between extractive activity and groundwater resources, an objective mathematical index has to be developed, with a multi-fold perspective.

First of all, it will determine an objective and reliable measure of the interaction. Indeed, it has been shown by \cite{Coll15} that experts in the field who were asked to measure the above defined interaction on a set of quarries, easily ended up with different conclusions. This is mainly because each expert has the subjective feeling that certain factors should have a larger impact, thus tends to overestimates them and underestimates others. 

Secondly, the mathematical index is capable to quantify in a single number what a time consuming 3D simulation could produce. Moreover, changes in a single parameter will require restart anew the simulation while the index immediately provides the new required information. Hence, the index allows to extrapolate the possible impact of future changes in the quarries or in the environment, as well as to provide a longitudinal evolution of the quarry.

Finally the index can be easily computed and used also by non-technical experts, such as public administration employees or decision makers. It must be a tool capable of providing an easy contextualization of a study site by clearly synthesizing the existing information. It will help to define how far the feasibility study of the exploited quarry should go into detailed hydrogeological investigations in order to ensure that quarrymen and water producers have quality water in sufficient quantity for their respective uses and to preserve the environment.

Many environmental impact evaluation methodologies have been developed for the mining activities (\citealt{WangYangXu06}), often relying on large interaction matrices (\citealt{MonjShahDegh09}).  These matrices, firstly introduced by \cite{Leop71} to assess the potential interactions between a project and environmental factors, have been developed and used in a large number of mining related studies (\citealt{AryaYousArdeDoul13}). For instance this methodology was applied to investigate the impact of the Yanjiao mine on health (\citealt{Szwi97}); the interaction between mining and water reservoir pollution in England (\citealt{JarvYoung00}); the effect of mining on surface water, groundwater and ground sinking (\citealt{BlodKuip02}); the environmental impact of mining with explosive (\citealt{Folc03}); the environmental, socio-demographic, economic and cultural impacts of mining activities (\citealt{MirmGholFattSeyeGhor09}). Recently, \citealt{AryaYousArdeDoul13} improved the interaction matrix method to fully integrate the interference of the mining operations with groundwater together with other environmental factors.

Despite the good results obtained with the interaction matrix method, the latter suffers form a main issue: the conjoint presence of numerical variables, namely quantities that enter in the model through dimensional number such as volumes, flows, ..., and categorical variables, namely involved quantities defined via abstract scales such as \textit{large}, \textit{medium}, \textit{small} or resulting from the transformation of numerical variables into representative intervals that is $A$, $B$, $C$. To handle the former cases it introduces a (usually large) set of criteria and subcriteria, for instance the \textit{net recharge} can be divided into \textit{Less than 5}, \textit{between 5 and 10}, \textit{between 10 and 18}, \textit{between 18 and 25} and \textit{more than 25} and such division is based on human decisions and thus subjected to the researcher appreciation of the problem. Moreover, such (sub)criteria are weighted using numerical values and this introduces a second issue. Indeed, a parameter whose value continuously changing allows its to switch from a (sub)category to the next one has often a non linear effect that cannot be completely described by a multiplicative weight. 

The methods we propose do not have such issues. Roughly speaking, the use of logical variables, describing the presence (\textit{True}) or the absence (\textit{False}) of a condition, allows to easily merge numerical and categorical variables and the weights are computed directly from the observations and thus intrinsically determined by the model. For example in the following we will use to characterize the hydrogeological context of a quarry, the following variables \textit{aquiclude formation}, \textit{aquitard formation}, \textit{aquifer formation} and \textit{carbonate aquifer formation} whose exclusive values can be \textit{True} or \textit{False}, denoting to which of the above classes the quarry under examination belongs to.

Alternative assessment methods include simulation (e.g. \citealt{EberBair90}), the use of environmental tracers (e.g. \citealt{DarlGoodRichWall10}), multi-criteria decision analysis such as the (fuzzy) analytical hierarchy process, input-output analysis, life cycle assessment and fuzzy set approaches (see \citealt{MonjShahDegh09}, for a review of those methods). 

To the best of the authors knowledge, a model that is simple yet able to correctly predict the level of interaction between quarries operations and groundwater is still missing in the literature. Using simple geological and hydrogeological parameters, we propose, in this work, two interaction indices to assess this interaction. The first is based on a discrete choice modelling approach while the second uses the artificial neural network methodology. Both interaction indices rely on the equation used in the assessment of natural hazards (\citealt{DaupProv03}), defined as
\begin{equation}
I = f(Quarry,~Groundwater),
\label{eq_dauphine}
\end{equation}
where the interaction $I$ corresponds to the environmental risk, which is a function of hazard and vulnerability parameters detailing respectively the quarry and the groundwater resources. Depending on the value of $I$, the quarry will present a \emph{low}, a \emph{medium}, a \emph{high}, or a \emph{very high} potential impact on the regional hydrogeology. Hence, $I$ determines the level of investigation of the feasibility study to undertake before considering any extension of the quarry (Table \ref{table_interaction_level}). What follows in the next sections aims to develop two ways to estimate this interaction level $I$.

\begin{table}[h!]
\centering
\begin{tabular}{|c|rcl|}
\hline
\textbf{Interaction index} & \multicolumn{3}{c|}{ \textbf{Feasibility study}}\\
\hline
\textit{low} & & & hydrogeology characterization \\ 
\textit{medium} & low & + & piezometric monitoring \\ 
\textit{high} & medium & + & steady state (static) mathematical model\\ 
\textit{very high} & high & + & transient state (dynamic) mathematical model\\ 
\hline
\end{tabular} 
\caption{Required level of investigation with respect to the interaction index}
\label{table_interaction_level}
\end{table}

The remaining part of this paper is organized as follows. The values of the interaction index as well as the parameters influencing it are firstly described in Section~\ref{sec_interaction}. Section~\ref{sec_discrete_choice} then illustrates a first interaction index based on the discrete choice methodology. This is followed in Section~\ref{sec_neural_network} by a second index relying on a neural network formulation. Their respective application on a particular quarry is illustrated and compared in Section~\ref{sec_application}. Concluding remarks and perspectives are presented in the last section.

\section{Interaction index and parameters}
\label{sec_interaction}

We have retained 6 main parameters, each classified into 4 modalities, to define the interaction index. These parameters are grouped in two distinct categories:
\begin{itemize}
  \item the geological ($G$),  hydrogeological ($H$) and piezometric ($Z$) contexts defining the hazard that a quarry represents and detailed in Table \ref{tab_param_1};
  \item the relative position of the quarry and the water catchments ($C$), the production of the catchments ($T$) and the potential quality of the groundwater ($L$) characterising the vulnerability of the groundwater resources and described in Table \ref{tab_param_2}.
  \end{itemize}
Each of the resulting 3327 physically feasible combinations of these parameters (out of a theoretical number of $4^6 = 4096$ possible combinations) determines one particular quarry site type (e.g. $G_4,H_3,Z_2,C_2,T_1,L_1$) referred to as a quarry type. See Appendix \ref{app_dataset} for a subset of the retained combinations.

The parameters and their modalities are fully detailed in Collier~et~al.~(2015). It should be noted that they were selected based on the authors own experience in the field and have been chosen (\citealt{Coll15}), among the huge number of possible other parameters, as the ones with the largest impact using dedicated 3D-hydrogeologic dynamical numerical simulations based on Visual
Modflow Flex (\citealt{Modf14}). Finally, this choice was also validated by several experts coming from water public operators (Aquale, Société Wallone Des Eaux), universities (University of Liege and University of Mons), the Public Service of Wallonia (SPW), quarry operators (Carmeuse, Lhoist and Carrières Unies de Porphyre) and consulting companies (Institut Scientifique de Service Public, Geolys).

The interaction index of a quarry type, ranging from low to very high (as defined in Table \ref{table_interaction_level}), is then a function of the above parameters, i.e. Equation \eqref{eq_dauphine} becomes:
\begin{equation}
I = f(G,H,Z,C,T,L).
\end{equation}
\label{eq_base}
How to formally define the relation between $I$ and the parameters is the topic of the next two sections.

\begin{table}[p]
\begin{tabularx}{\textwidth}{|>{\setlength\hsize{0.125\hsize}\centering}X@{}|>{\setlength\hsize{1.875\hsize}}X|}
  \hline
  \multicolumn{2}{|c|}{\textbf{Geological context}}\\
  
  \mc{2}{Characterizes the lithology and extension of the geological formation exploited in the quarry and those of the neighbouring geological formations that will govern the groundwater flow directions.}\\
  \hline
  $G_1$ & completely isolated by other formations with low permeability\\
  $G_2$ & limited extension and partly compartmentalized\\
  $G_3$ & local extension\\
  $G_4$ & regional extension\\
  \hline
  \multicolumn{2}{|c|}{\textbf{Hydrogeological context}}\\

  \mc{2}{Refers to the combinations of geological formations according to their hydrodynamic characteristics.}\\
  \hline
  $H_1$ & aquiclude formation\\
  $H_2$ & aquitard formation\\
  $H_3$ & aquifer formation\\
  $H_4$ & carbonate aquifer formation\\
  \hline
  \multicolumn{2}{|c|}{\textbf{Piezometric context: altimetric level of the quarry floor}}\\

  \mc{2}{Characterizes the relative position between the quarry pit bottom and the groundwater piezometric level.}\\
  \hline
  $Z_1$ & higher than the piezometric level of the water table\\
  $Z_2$ & lower than the piezometric level of the water table but higher than the river thalweg which is the regional base level\\
  $Z_3$ & lower than the piezometric level of the water table and the altimetric level of the river thalweg which is the regional base level\\
  $Z_4$ & lower than the piezometric level of the water table and the altimetric level of the river thalweg which is not the regional level any more (the river is perched)\\
  \hline
\end{tabularx}
\caption{Hazard (quarry) parameters}
\label{tab_param_1}
~\\~\\
\begin{tabularx}{\textwidth}{|>{\setlength\hsize{0.125\hsize}\centering}X@{}|>{\setlength\hsize{1.875\hsize}}X|}
  \hline
  \multicolumn{2}{|c|}{\textbf{Relative position of the quarry and the water catchments}}\\
  \mc{2}{Catchments (well, spring, gallery, etc.) for public distribution of drinking water are threatened by various sources of pollution. Closer a quarry gets to the catchment, greater its impact may be important. Consequently, 4 successive perimeters, within which the activities and facilities are regulated, are set up around the catchment based on the velocity of groundwater (transfert time).}\\
  \hline
  $C_1$ & outside the drainage zone of a catchment\\
  $C_2$ & in the drainage zone of a catchment\\
  $C_3$ & in the distant prevention area of a catchment (50 days of delay in case of aquifer contamination)\\
  $C_4$ & in the close prevention area of a catchment (24 hours of delay in case of aquifer contamination)\\
  \hline   
  \multicolumn{2}{|c|}{\textbf{Production of the catchments}}\\
  \mc{2}{Volume exploited in catchments for public distribution in the hydrogeological formation near the quarry.}\\
  \hline
  $T_1$ & lower than $2~m^3 / h$\\
  $T_2$ & between 2 and $10~m^3 / h$\\
  $T_3$ & between 10 and $30~m^3 / h$\\
  $T_4$ & greater than $30~m^3 /h$\\
  \hline   
  \multicolumn{2}{|c|}{\textbf{Potential quality of the catchments}}\\
  \mc{2}{Quality and the potability of the groundwater.}\\
  \hline
  $L_1$ & poor quality\\
  $L_2$ & water potabilisable with minor treatment\\
  $L_3$ & good quality water\\
  $L_4$ & water of exceptional quality (mineral water)\\
  \hline
\end{tabularx}
\caption{Vulnerability (groundwater) parameters}
\label{tab_param_2}
\end{table}

\section{A Logit discrete choice model-based index}
\label{sec_discrete_choice}

This section details a first interaction indicator relying on the discrete choice methodology. A discrete choice model aims at describing, explaining and predicting the choice made by an entity amongst a set of alternatives $\mathcal{A}$  (\citealt{BenALerm85}). This choice set must be exhaustive and contain a finite number of mutually exclusive alternatives.

Discrete choice models rely on the utility maximisation theory, i.e. they assume that the entity will always opt for the alternative that maximizes its utility (or benefit). Let us denote by $U_{jn}$ the utility that the entity $n$ associates with the alternative $j$, $\forall~j \in \mathcal{A}$. Then $n$ will choose the alternative $i$ if 
$$
U_{in} > U_{jn} \quad \forall j \neq i.
$$ 
The entity choice depends on many factors, some of those may remain unknown for an external observer\footnote{For instance the choice of a drink (an alternative) by an individual (the entity) in a pub depends on observed characteristics (e.g. time of the day, gender of the individual, age) and personal preferences of the individual (e.g. its mood, health condition, personal taste) which are unknown.}. The utility that an entity obtains from choosing an alternative is therefore decomposed into two parts: 
$$
U_{in} = V_{in} + \epsilon_{in},
$$
where
\begin{itemize}
  \item $V_{in}$ is the observed representative (or systematic) utility perceived by the entity $n$ for the alternative $i$;
  \item $\epsilon_{in}$ denotes the hidden (or random) component involved in the choice process.
\end{itemize}
The probability for an entity $n$ to retain alternative $i$ is then simply
\begin{eqnarray}
  P_{in} & = & P(U_{in} > U_{jn}, \quad \forall j \neq i)\\
             & = & P(\epsilon_{jn} - \epsilon_{in} < V_{in} - V_{jn}, \quad \forall j \neq i).
\end{eqnarray}
These equations highlight two important features of discrete choice models: only the differences in utility matter\footnote{Not the absolute value of the utilities.} and the overall scale of the utilities is irrelevant\footnote{Multiplying each utility by any arbitrary strictly positive factor $\alpha$ does not change the ordering of the utilities.}. The former implies that one of the alternative is associated with an utility set to $0$ and is referred to as the \emph{base} alternative. Generally the representative utility can be put into the form (linear model)
\begin{equation}
V_{in} = \beta_i^{*T} x_{in},
\label{eq_beta_x}
\end{equation}
where $x_{in}$ and $\beta_i^*$  are two vectors, the former detailing the observed variables for the entity $n$ with respect to the alternative $i$, and the latter containing the corresponding coefficients referred to as the model parameters.
Note that these two vectors, collected for all alternatives $i$ in $\mathcal{A}$, give vectors $x_n$ and $\beta^*$, respectively.

Different specifications of the random component $\epsilon_{in}$ lead to different models. The Logit model, used in this work, is derived by assuming that these terms are independent (i.e. no correlation between the alternatives) and follows an identical standard Gumbell distribution whose cumulative distribution function is defined by:
\begin{equation}
  F(\epsilon_{in}) = e^{-e^{-\epsilon_{in}}},
\end{equation}
where $e$ is the Euler's constant. As a result, it can be shown that the probability associated with each alternative has the form (\citealt{BenALerm85}):
\begin{equation}
P_{in} = \frac{e^{V_{in}}}{\sum_k e^{V_{kn}}}.
\label{eq_proba}
\end{equation} 
The vector of coefficients $\hat{\beta}$ estimating $\beta^*$ in Equation \eqref{eq_beta_x} for this Logit model is then obtained by the maximum likelihood approach, i.e.:
\begin{equation}
\hat{\beta} = \argmax_{\beta}  \mathcal{L}(\beta) = \sum_n \sum_i y_{in} \left( \beta^T_i x_{in} - \ln \sum_j e^{\beta^T_j x_{jn} } \ \right),
\end{equation}
where the indicator variable $y_{in} = 1$ if $n$ effectively chooses $i$, $0$ otherwise. A goodness of fit measure of the model is given by the value of the index of likelihood ratio $\rho^2 \in [0,1]$ defined as:
\begin{equation}
  \rho^2 = 1 - \frac{\mathcal{L}(\hat{\beta})}{\mathcal{L}(0)}.
\end{equation}
If the fitted model is able to perfectly reproduce the entities choices then $\rho^2 = 1$. Conversely $\rho^2 = 0$ if the model is no better than a model where all the coefficients are set to $0$ and thus the choices are dictated by random decisions. Note that since the value of $\rho^2$ increases with the number of parameters, the adjusted $\bar{\rho}^2$ has been introduced to compare models with different number of parameters and defined by: 
\begin{equation}
  \bar{\rho}^2 = 1 - \frac{\mathcal{L}(\hat{\beta}) - K}{\mathcal{L}(0)},
\end{equation}
where $K$ is the number of estimated parameters. 

We refer the interested reader to \cite{BenALerm85}, and \cite{Trai09}, for an extensive description of the discrete choice methodology.

\subsection{The discrete choice Logit index}

Let us first define the choice set $\mathcal{A}$ available to every quarry $n$ (i.e. the entity in this context). As mentioned previously, the alternatives correspond to the four levels of the interaction index presented in Table \ref{table_interaction_level}, resulting in the following choice set:
$$
\mathcal{A} = \{low,~medium,~high,~very~high\},
$$
where \emph{low} has been chosen to be the base alternative. The utility associated with each alternative will be respectively denoted by $V_{l}$, $V_{m}$, $V_{h}$, and $V_{v}$ (we omit the quarry index $n$ for clarity).

One can easily observe that the alternatives in $\mathcal{A}$ are independent; furthermore it may be impossible to fully characterize  a quarry (for instance, due to cost reasons, some areas of a quarry might be left uninvestigated) resulting in hidden attributes for the observer. Hence, the assumption of the Logit discrete choice model are met and it can be applied in our framework.

The utility functions of the complete model encompassing every vulnerability and hazard parameters (detailed in Tables \ref{tab_param_1} and \ref{tab_param_2}) are expressed as:
\begin{eqnarray}
  V_{l} & = & 0, \label{eq_v_low}\\
  V_{m} & = & \beta^m + \sum_{k=1}^4 \left(\beta_{g,k}^m G_k + \beta_{h,k}^m H_k + \beta_{z,k}^m Z_k + \beta_{c,k}^m C_k + \beta_{t,k}^m T_k + \beta_{l,k}^m L_k\right),\\
  V_{h} & = & \beta^h + \sum_{k=1}^4 \left(\beta_{g,k}^h G_k + \beta_{h,k}^h H_k + \beta_{z,k}^h Z_k + \beta_{c,k}^h C_k + \beta_{t,k}^h T_k + \beta_{l,k}^h L_k\right),\\
  V_{v} & = & \beta^v + \sum_{k=1}^4 \left(\beta_{g,k}^v G_k + \beta_{h,k}^v H_k + \beta_{z,k}^v Z_k + \beta_{c,k}^v C_k + \beta_{t,k}^v T_k + \beta_{l,k}^v L_k\right),
\end{eqnarray}
resulting in a vector $\beta$ of unknown parameters to be estimated. Equation \eqref{eq_v_low} is a consequence of \emph{low} being the base alternative. The terms $\beta^i$, $i \in \{m, h,v\}$ correspond to the {\em alternative specific constants}, capturing the average effects of the hidden (errors) components $\epsilon_i$. The quarry parameters are coded as dummy variables with values being 0 or 1 depending on the quarry characteristics, for instance
\begin{equation}
G_i = 1 \Leftrightarrow G_j = 0 \quad \forall j \neq i,
\end{equation}
and similarly for $\{H,Z,C,T,L\}$, i.e. the modalities of each parameter are mutually exclusive. The probabilities of each level $P_l$, $P_m$, $P_h$ and $P_v$ are finally derived from Equation \eqref{eq_proba}.

The maximum likelihood estimation of the model parameters, i.e. the components of $\beta$, has been performed using the BIOGEME software (\citealt{Bier03}) on the whole theoretical data set described in Section \ref{sec_interaction} and Appendix \ref{app_dataset}. The summary statistics of this estimation process are detailed in Table \ref{tab_sum_stat_1} and indicate that the estimation process converged to a model which has satisfactory performance with a final log-likelihood of $\mathcal{L}(\hat{\beta})$ higher than $\mathcal{L}(0)$, resulting in a $\rho^2$ of 0.623 and a $\bar{\rho}^2$ of 0.611.

\begin{table}[h!]
\centering
\begin{tabular}{|lc|}
\hline
\multicolumn{2}{|c|}{\textbf{Summary statistics}}\\
\hline
Model & Logit\\
Number of estimated parameters & 57\\
Number of  observations & 3327\\
Null log-likelihood $\mathcal{L}(0)$ & -4612.201\\
Final log-likelihood $\mathcal{L}(\hat{\beta})$ & -1738.991\\
$\rho^2$ & 0.623\\
$\bar{\rho}^2$ & 0.611\\
Diagnostic & Convergence reached\\
Variance-covariance & from analytical Hessian\\
\hline
\end{tabular}
\caption{Summary statistics of the complete model}
\label{tab_sum_stat_1}
\end{table}

The value of the estimated coefficients are listed in Appendix \ref{app_first_logit} together with their respective standard error (derived from the variance-covariance matrix determined by BIOGEME), $t$-statistic ($t$) and associated $p-value$ ($p$) which were used to test the statistical significance of the coefficients. In this work, we retain a confidence level of $95\%$, meaning that a coefficient is not statistically different from $0$ if $|t| \leq 1.96$ or equivalently if $p \geq 0.05$.

It can then be observed that nine of the parameters are not statistically significant. This issue leads us to fit a second model wherein insignificant coefficients are not estimated any more but are instead fixed to $0$.  After the convergence of the estimation process, every estimated coefficient of the second model are now statistically significant (see Appendix \ref{app_second_logit}). The final values of the parameters, both fixed and estimated, are detailed in Table \ref{tab_estimates}. Summary statistics of this simplified model can be found in Table \ref{tab_sum_stat_2}. It can be observed that, despite that the number of parameters to estimate has been reduced, the $\bar{\rho}^2$ and  $\mathcal{L}(\hat{\beta})$ values of $0.609$ and $-1754.804$ are very close to the ones associated with the complete model (respectively $0.611$ and $-1738.991$). As a result, we will retain this second model as the discrete choice-based interaction index.

\begin{table}[h!]
  \centering
  \begin{tabular}{|c|c|r|r|r|r|r|r|r|}
    \cline{2-9}
    
 \multicolumn{1}{c|}{} &  \multirow{2}{*}{$k$} & \multirow{2}{*}{$\hat{\beta}^i_{g,k}$} & \multirow{2}{*}{$\hat{\beta}^i_{h,k}$} & \multirow{2}{*}{$\hat{\beta}^i_{p,k}$} & \multirow{2}{*}{$\hat{\beta}^i_{c,k}$} & \multirow{2}{*}{$\hat{\beta}^i_{t,k}$} & \multirow{2}{*}{$\hat{\beta}^i_{l,k}$} & \multicolumn{1}{c|}{\multirow{2}{*}{$\hat{\beta}^i$}}\\
   
    \multicolumn{1}{c|}{}  & &  &  &  &  & &  &\\
    
    
    \hline
    \multirow{4}{*}{$V_m$} & 1 &  0.00 & 0.00 &  0.00 & 0.00 & 0.00 &  0.00 & \multirow{4}{*}{-9.42} \\
                                              & 2 & -0.38 & 0.00 &  2.21 & 3.26 & 0.45 &  4.08 &\\
                                              & 3 & -0.39 & 2.97 &  4.65 & 5.35 & 1.69 & 4.64 &\\
                                              & 4 & -0.44 & 2.97 &  4.59 & 5.34 & 4.50 & 4.78 &\\        
    \hline
    \multirow{4}{*}{$V_h$} & 1 & 0.00 &  0.78 & -5.35 & -6.92 & -3.57 & -7.20 & \multirow{4}{*}{-1.97}\\
                                             & 2 & 0.00 &  0.00 &  0.00 &  0.00 & 0.00 &    0.00 &\\
                                             & 3 & 0.00 &  5.11 &  3.88 &  4.99 & 2.97 &    1.91 &\\
                                             & 4 & 0.00 &  5.11&   3.84 &  5.35 & 6.47 &    3.24 &\\        
    \hline
    \multirow{4}{*}{$V_v$} & 1 & 0.00 & -7.36 & -16.90 & -23.50 & -13.00 & -13.90 & \multirow{4}{*}{24.20}\\
                                             & 2 & 0.00 & -8.96 &   -7.69 &   -8.48 & -6.16 & -3.66 &\\
                                             & 3 & 0.00 &  0.00 &    0.00 &    0.00 &  0.00 & 0.00 &\\
                                             & 4 & 0.00 &  0.00 &    0.00 &    0.78 & 2.53 & 3.30 &\\                 
    \hline                                                                                                          
  \end{tabular}
  \caption{Coefficient estimates of the final Logit-based indicator. Note that the level $low$ is omitted since it is the base alternative and $V_l = 0$}
  \label{tab_estimates}
\end{table}

\begin{table}[h!]
\centering
\begin{tabular}{|lc|}
\hline
\multicolumn{2}{|c|}{\textbf{Summary statistics}}\\
\hline
Model & Logit\\
Number of estimated parameters & 48\\
Number of  observations & 3327\\
Null log-likelihood $\mathcal{L}(0)$ & -4612.201\\
Final log-likelihood $\mathcal{L}(\hat{\beta})$ & -1754.804\\
$\rho^2$  &   $0.620$ \\
$\bar{\rho}^2$ &    $0.609$ \\
Diagnostic & Convergence reached\\
Variance-covariance & from analytical Hessian\\
\hline
\end{tabular}
\caption{Summary statistics of the model keeping only significant parameters}
\label{tab_sum_stat_2}
\end{table}

\section{Neural network-based index}
\label{sec_neural_network}

An artificial neural network is a supervised machine learning algorithm inspired by the biological neural networks: artificial neurons sense stimuli (inputs) from other neurons, analyse the informations acquired and pass the result to the following neurons if the stimuli is strong enough (i.e. above a given activation threshold). Typically artificial neural networks are organized in interconnected layers of neurons referred to as nodes. The feed-forward neural network architecture retained in this work is characterized by one input layer, at least one hidden processing layer and one output layer. Each of those layers contains at least one node. The information is then passed from one layer to the next one. Due to the number of layers and the complexity of the nodes interconnections, this approach is often compared to a black box. It has been shown that a neural network is able to approximate any function (universality property). Hence, this approach seems appropriate to design an interaction index. An extensive introduction to artificial neural networks, which has been applied in countless applications, can be found in \cite{Ripl96}, and \cite{Krie07}.

The nodes belonging to two successive layers are connected by a weighted link carrying the stimuli through the network if the origin node has been activated. Formally, the links between each pair of nodes $(o_i^l, o_j^{l+1})$ belonging to two successive layers $l$ and $l+1$ are characterized by a weight $w_{ij}^l$. The inputs are then passed through one hidden layer to the next one using the activation thresholds in combination with the logistic transfert function. The state of a node $o_j^{l+1}$ given the state of the nodes in the previous layer is then defined by:
\begin{equation}
f(o_j^{l+1}) = \max \left\{ 0,\sigma(\vec{o}^{l})\right\},
\end{equation}
where $\sigma(\vec{o}^l)= ({1 + e^{- \sum_i w_{ij}^l f(o_i^l) }})^{-1} - \theta_j^{l+1}$, $\vec{o}^{l}=(o_1^l,\dots,o_{n_l}^l)$ is the vector of the nodes states in the layer $l$ and $\theta_j^{l+1}$ is the activation thresholds of the node $o_j^{l+1}$ . Note that the nodes belonging to the input and output layers do not have an activation function.

The vector of inputs and outputs of this neural network-based index is identical to the ones used by the previous discrete choice-based index, i.e. the input layer has 24 nodes representing the quarry parameters coded again as dummy variables and the output layer contains four nodes referring to the four possible levels. Finally, the following decision rule is applied to determine the interaction level:
\begin{equation}
I = \argmax_{l,m,h,v} \left\{ O_l, O_m, O_h, O_v \right\}
\end{equation}
where $O_l$, $O_m$, $O_h$ and $O_v$ respectively denote the value of the output node associated with the high, medium, high and very high level of interaction. It should be noted that the output layer is normalized: each output node $O_i\in [0,1]$ and $\sum_i O_i = 1$, $i \in \{l, m, h, v \}$.

The neural network is fully defined by the number of hidden layers, the number of nodes in these layers, the weights of the links interconnecting the nodes and the activation thresholds. As stated previously, we will consider only one hidden layer in this work. Several designs have then been tested by varying the number of nodes in this hidden layer from 1 to 17 in order to determine the optimal number of nodes in this layer.  The goodness of fit measure of  performance for each tested network is simply given by the percentage of accurate prediction of the interaction index. 

The coefficients of the neural network are optimised by a resilient back-propagation with weight backtracking learning algorithm (\citealt{RiedBrau93}) performed by the R package \emph{neuralnet} (\citealt{FritGuenSuli12,R16}). The number of repetitions has been set to 10 and the error function minimised by the training algorithm is the sum of square errors between the predicted and expected level of interaction for each quarry in the training set. The training process has been performed on a $75\%$ simple random sample extracted from the theoretical dataset. 

\begin{figure}[p!]
  \centering
  \includegraphics[scale=0.6]{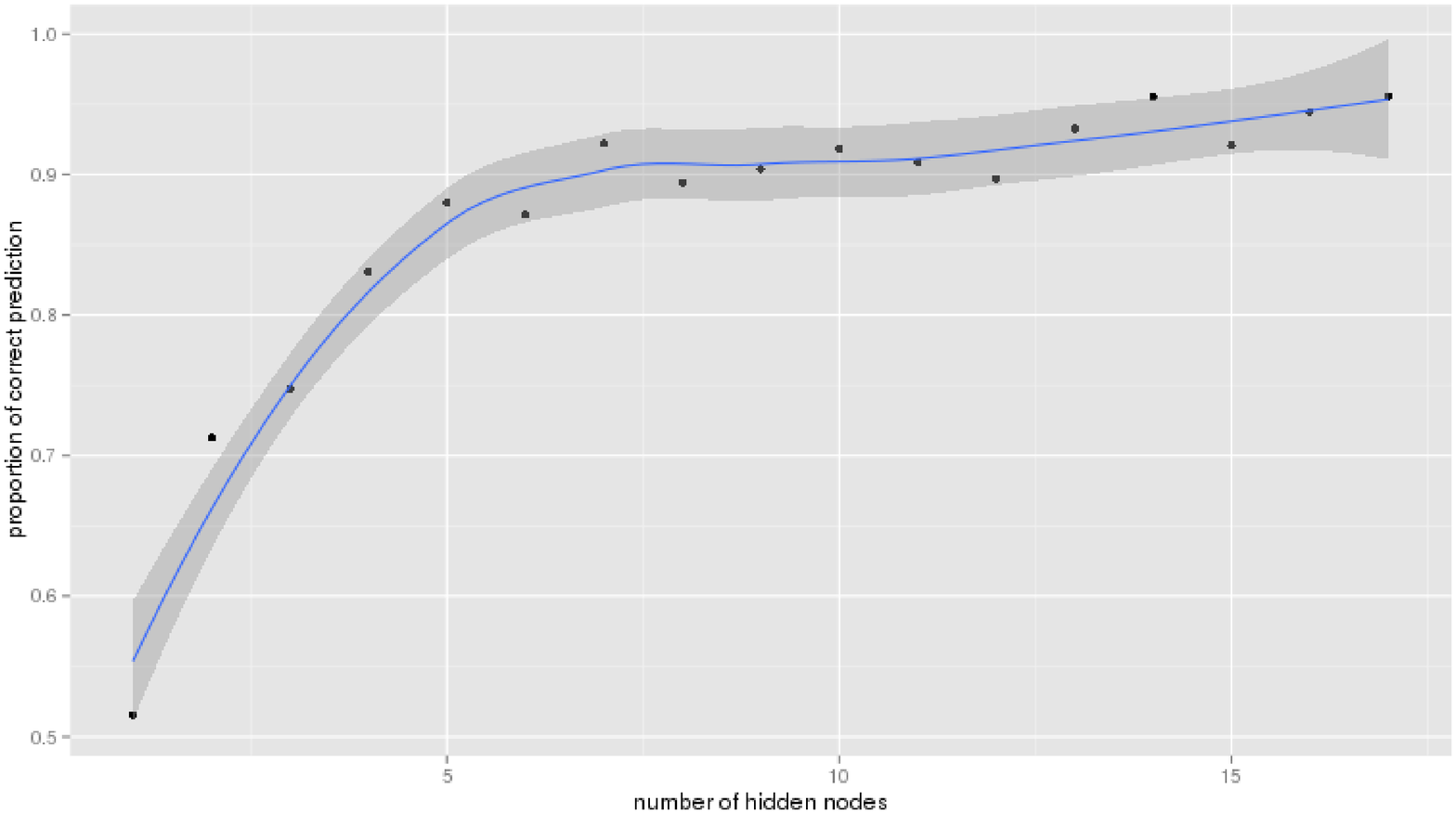}
  \caption{Proportion of correct prediction with respect to the number of nodes in the hidden layer. The solid line corresponds to the local regression of the results along with its standard error defining the confidence limit of the regression.}
  \label{fig_nnet_predictions}
  ~\\~\\
  \centering
  \includegraphics[scale=0.40]{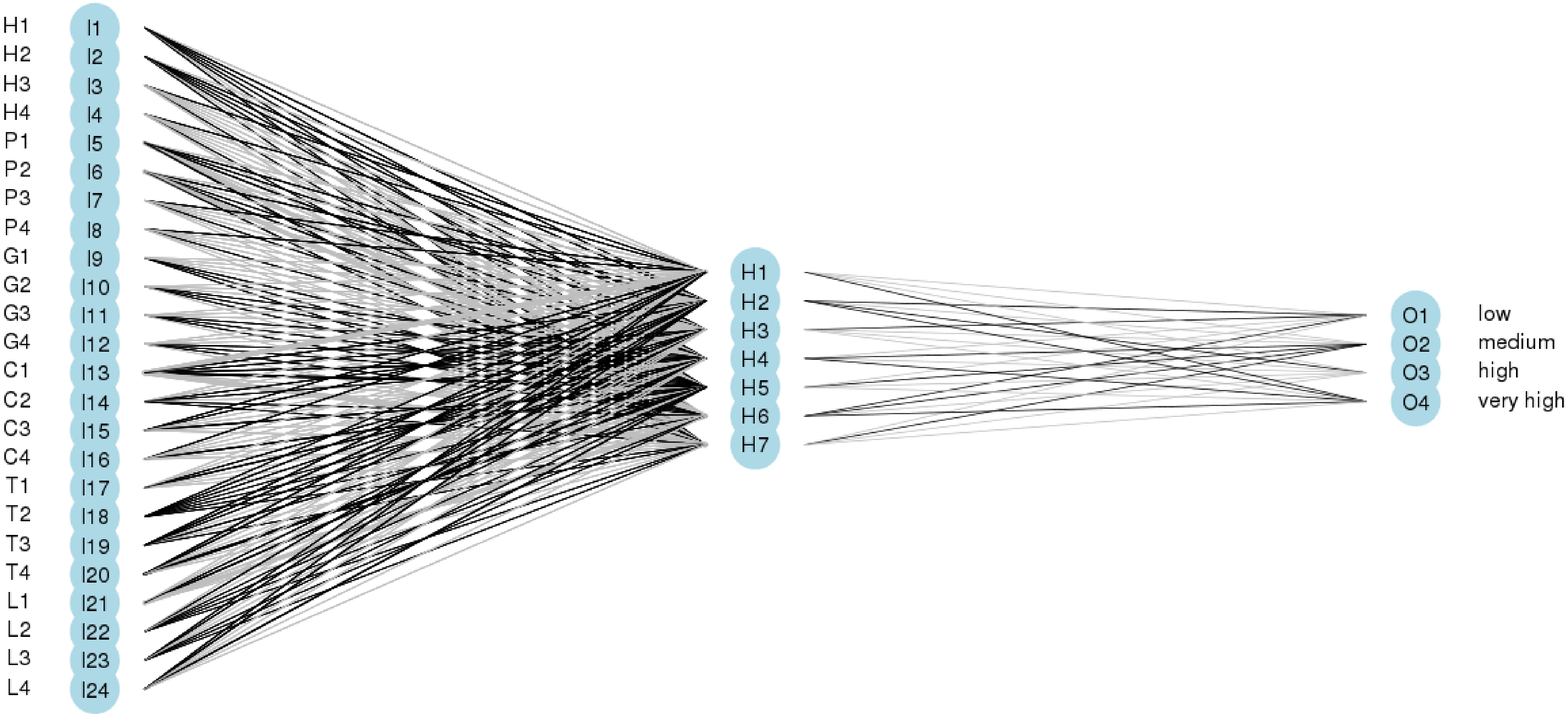}
  \caption{The retained feed-forward neural network architecture with 7 nodes in the hidden layer. Percentage of correct prediction $= 92.19\%$. The decision rule is to select the output node associated with the highest value. The weights and the activation thresholds are detailed in Appendix \ref{app_neural_net}. A black link indicates that the associated weight is positive, while a grey link indicates a negative one.} 
  \label{fig_nn}
\end{figure}

The performance of the resulting trained network has then been assessed on the remaining $25\%$ of the data (which was not used in the training step). The results of the validation experiments are shown in Figure \ref{fig_nnet_predictions}. One can easily observe that the percentage of correct prediction rapidly increases from 1 hidden node ($51.56\%$) to 7 ($92.19\%$). Beyond that threshold, the impact of additional nodes is lower but remains positive. In this work we will then select a neural network with 7 nodes in the inner layer whose architecture is illustrated in Figure~\ref{fig_nn}. The high rate of correct predictions for quarries that were not present in the training data tends to indicate that the neural network is capable to assess the interaction level of new quarries.

Once the indices proposed in this Section and the previous one for predicting the interaction level between the extractive activity and groundwater resources have been obtained, one can illustrate how to use them on a particular quarry. This is the topic of the next Section.

\section{Application on the Boverie quarry and the Tridaine gallery}
\label{sec_application}

\begin{figure}[h!]
  \centering
  \includegraphics[scale=0.38]{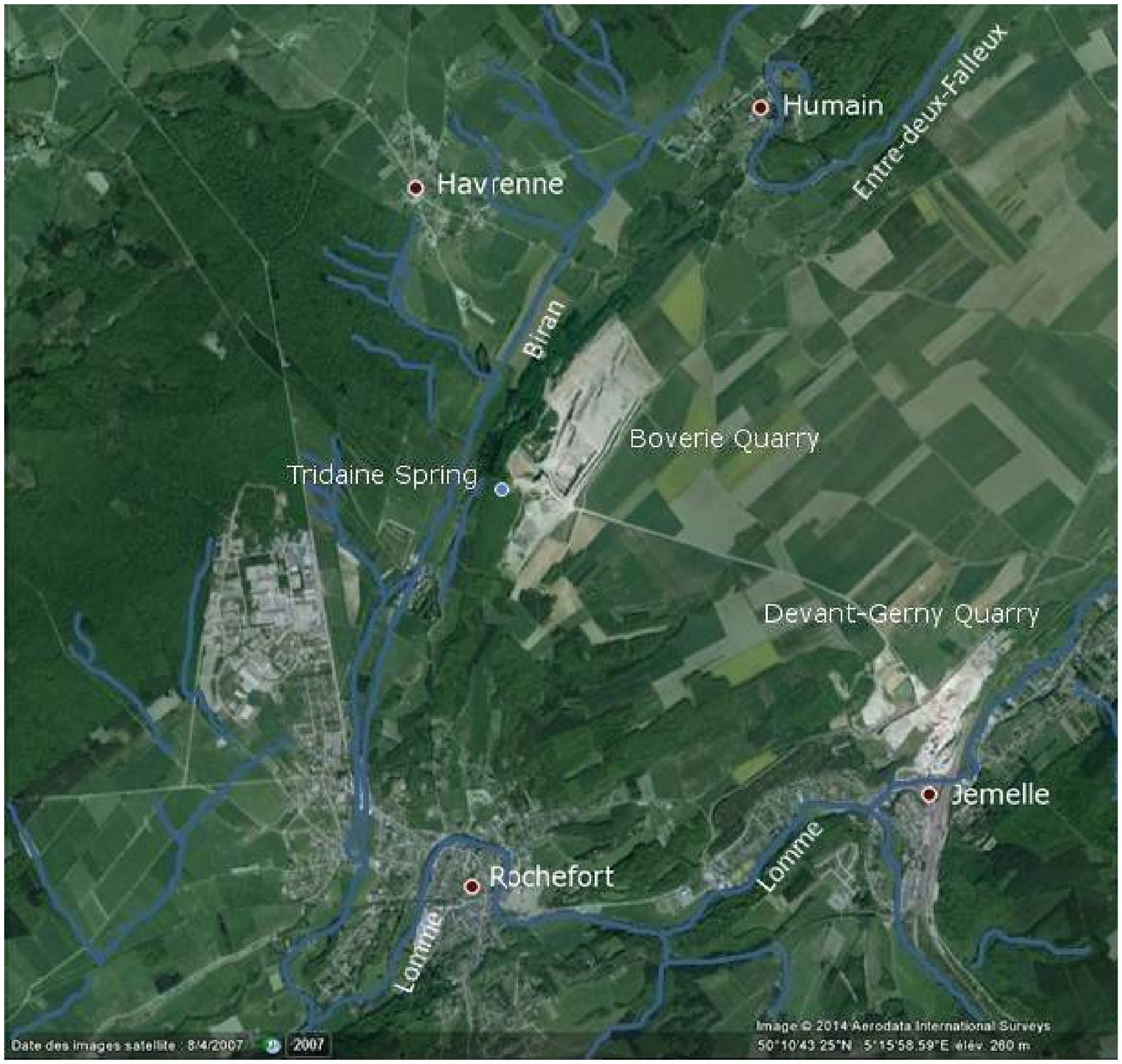}
  \caption{Quarry of Boverie and the Tridaine gallery (Google Earth)}
  \label{fig_quarry}
\end{figure}

We now demonstrate the application of the Logit-based and the neural network indices on the Boverie quarry and the Tridaine spring located on the Gerny plateau, a limestone massif of an area of 1.500 hectares, extending at the north-east of the city of Rochefort in the Walloon Region of Belgium (see Figure \ref{fig_quarry}). The Frasnian limestones exploited by the Lhoist Group in the Boverie quarry and the groundwater of this aquifer pumped from the Tridaine gallery are in the center of important economic issues. Indeed, these limestones have a high purity as they are used in the production of lime. The Tridaine spring, which results from an overflow of the groundwater contained within the Frasnian limestones, supplies the Saint-Remy’s Abbey and provides drinking water for the agglomeration of Rochefort. The interested reader can found an extensive study of this quarry in Collier et al. (2015).

The quarry and the groundwater are currently characterized by the following attributes: $G_2, H_4, Z_1, C_4, T_4$ and $L_3$. In order to assess the current situation of the studied quarry, the interaction indices defined in the previous sections are now applied with those attributes values and their results are compared.
\begin{itemize}
\item The discrete choice-based index produces the following probabilities:
    $$
    P_l \leq 0.01\%, \quad P_m = 1.50\%, \quad P_h = 66.4\%, \quad P_v = 32.1\%.
    $$
    This indicates that the quarry is very likely to have a \emph{high} interaction level due to the probability associated with this level. Nevertheless, even tough the probability of a \emph{very high} level is lower it is still significant. Hence, this possibility should not be totally excluded.
\item The neural network-based index returns the following values for the output nodes:
  $$
  O_l = 0.002, \quad O_m = 0.001, \quad O_h = 0.037, \quad O_v = 0.959,
  $$
  which indicates that there is a \emph{very high} level of interaction.
\end{itemize}
The combination of the two indices tends to suggest a very high level of interaction which matches with the true level interaction for that category of quarry in the data used to train both indicators. It is then safe to assume that a complete mathematical hydrodynamic and geographic modelling of the quarry is necessary (see Table \ref{table_interaction_level}). 

We now illustrate how the two indices can also be used to predict interactions levels resulting from future quarry extensions. Currently the limestone lenses are exploited at an altimetric level of 220 $m$, which is higher than the groundwater table (211.5 $m$). Nevertheless, it is planned to deepen the exploitation down to a level of 160 $m$, resulting into the piezometric context switching from the category $Z_1$ to $Z_4$. The updated values of the interaction indices are then given below.
\begin{itemize}
  \item The probabilities of the discrete choice-based index become:
    $$
    P_l \leq 0.01\%, \quad P_m \leq 0.01\%, \quad P_h \leq 0.01\%, \quad P_v = 99.9\%,
    $$
    indicating that the interaction level is certainly \emph{very high}.
  \item This level is confirmed by the outputs of the neural network-based indicator:
    $$
    O_l = 0.002, \quad O_m = 0.001, \quad O_h = 0.035, \quad O_v = 0.961.
    $$
\end{itemize}
In this scenario both indices agree on the conclusion that reaching a level of 160 $m$ will result on a \emph{very high} interaction level as one would have expected. Accordingly, the operator will then be suggested again to perform a study up to mathematical modelling of the transient states.

\section{Conclusion and perspective}
\label{sec_conclusions}

In this work we designed two interaction indices between extractive activity and groundwater resources using two different methodologies, namely the discrete choice models and artificial neural networks. The proposed methods overcome the issues usually associated with the existing matrix methods, namely the large number of required parameters based on human decisions and the use of numerical values instead of categories. An other difference is that each method detailed in this paper is not only descriptive, but also predictive. Indeed, unlike the matrix methods, they are able to determine both the current value of interaction index, and the future (or past) ones when the configuration of the quarry changes over time. These methods have been implemented in the R package \emph{quarrint} (\citealt{BartCarlCollHallMoriSart16}).

It has been shown that the proposed indices produce satisfactory results and are complementary. A notable finding was that the artificial neural network seems to perform better than the Logit discrete choice approach in terms of predicting the right level of interaction in our application. Nevertheless, the former approach lacks the ability to statistically test the significance (and relevance) of the input parameters compared to the latter. Also, an other major difference between the two indices is the probabilistic nature of the Logit-based index, while the neural network consists of a deterministic computational black box. 

Depending on the value of the interaction indices, which can be low, medium, high or very high, the quarry operator will be advised to conduct an appropriate feasibility study. These indices have been successfully applied on a particular quarry to demonstrate their use and complementarity. 

Future works include investigating the applicability and extension of this work to other type of mining activities, as well as testing other indices. These indices can be based for instance on alternative neural network design, logistic regression, decision trees, other discrete choice model relying on different utility functions, etc.

\section*{Acknowledgments}
The work of T. Carletti, M. Moriamé, and A. Sartenaer presents research results of the Belgian Network DYSCO (Dynamical Systems, Control, and Optimization), funded by the Interuniversity Attraction Poles Programme, initiated by the Belgian State, Science Policy Office. This reseach is part of a project with the support and funding of the Public Service of Wallonia, in collaboration with the FEDeration of the Extractive Industry of Belgium (FEDIEX) and the professional union of the water public operators of Wallonia (Aquawal). The authors also gratefully acknowledge the support of NVIDIA Corporation with the donation of the GTX980 used for this research.

\begingroup
\parindent 0pt
\def\enotesize{\normalsize}
\theendnotes
\endgroup

\section*{Conflict of Interest}
The authors declare that they have no conflict of interest.

\nocite{CollBarthCarlMoriSartHall15}
\bibliographystyle{spbasic}	
\bibliography{refs}

\clearpage
\newpage
\appendix
\section{Appendix}

\subsection{Training data set}
\label{app_dataset}

\begin{table}[h]
\centering
\begin{tabular}{|c|c|c|c|c|c|c|c|}
\hline
\textbf{Id} & \textbf{Interaction index} & \textbf{H} & \textbf{Z} & \textbf{G} & \textbf{C} & \textbf{T} & \textbf{L}\\
\hline
1 & low & 1 & 1 & 1 & 1 & 1 & 1\\
2 & low &1 & 1 & 1 & 2 & 1 & 1\\
3 & low & 1 & 1 & 1 & 3 & 1 & 1\\
4 & low & 1 & 1 & 1 & 4 & 1 & 1\\
\vdots & \vdots & \vdots & \vdots & \vdots & \vdots & \vdots & \vdots\\
1511 & medium & 4 & 3 & 2 & 1 & 1 & 1\\
1512 & medium & 4 & 3 & 2 & 2 & 1 & 1\\
1513 & medium & 4 & 3 & 2 & 3 & 1 & 1\\
1514 & medium & 4 & 3 & 2 & 4 & 1 & 1\\
\vdots & \vdots & \vdots & \vdots & \vdots & \vdots & \vdots & \vdots\\
2198 & high & 4 & 1 & 1 & 3 & 2 & 4\\
2199 & high & 4 & 1 & 1 & 4 & 2 & 4\\
2200 & high & 4 & 1 & 1 & 3 & 3 & 2\\
2201 & high & 4 & 1 & 1 & 4 & 3 & 2\\
\vdots & \vdots & \vdots & \vdots & \vdots & \vdots & \vdots & \vdots\\
3324 & very high & 4 & 4 & 4 & 4 & 4 & 3\\
3325 & very high & 4 & 4 & 4 & 2 & 4 & 4\\
3326 & very high & 4 & 4 & 4 & 3 & 4 & 4\\
3327 & very high & 4 & 4 & 4 & 4 & 4 & 4\\
\hline
\end{tabular}
\caption{Theoritical physically feasible combinations of the quarry parameters and their associated interaction level. Example of non-feasible combinations include ($H_1, Z_1, G_1, C_3, T_1, L_2$), ($H_1, Z_4, G_1, C_4, T_3, L_4$) and ($H_2, Z_4, G_3, C_2, T_4, L_4$)}
\label{tab_data}
\end{table}

\subsection{Complete Logit model}
\label{app_first_logit}

\begin{table}[h!]
\centering
\begin{tabular}{rlr@{.}lr@{.}lr@{.}lr@{.}l}
         &                       &   \multicolumn{2}{l}{}    & \multicolumn{2}{l}{\textbf{Robust}}  &     \multicolumn{4}{l}{}   \\
\textbf{Parameter} &                       &   \multicolumn{2}{l}{\textbf{Coeff.}}      & \multicolumn{2}{l}{\textbf{Asympt.}}  &     \multicolumn{4}{l}{}   \\
\textbf{number} &  \textbf{Description}                     &   \multicolumn{2}{l}{\textbf{estimate}}      & \multicolumn{2}{l}{\textbf{std. error}}  &   \multicolumn{2}{l}{\textbf{$t$-stat}}  &   \multicolumn{2}{l}{\textbf{$p$-value}}   \\

\hline
1 & ASC2 & -9&30 & 0&511 & -18&22 & 0&00 \\
2 & ASC3 & -2&09 & 0&361 & -5&78 & 0&00 \\
3 & ASC4 & 24&6 & 0&846 & 29&12 & 0&00 \\
4 & BETAC1\_O & -6&94 & 0&380 & -18&25 & 0&00 \\
5 & BETAC1\_R & -23&9 & 1&31 & -18&30 & 0&00 \\
6 & BETAC2\_J & 3&29 & 0&247 & 13&34 & 0&00 \\
7 & BETAC2\_R & -8&62 & 0&428 & -20&15 & 0&00 \\
8 & BETAC3\_J & 5&40 & 0&330 & 16&40 & 0&00 \\
9 & BETAC3\_O & 5&00 & 0&308 & 16&24 & 0&00 \\
10 & BETAC4\_J & 5&40 & 0&336 & 16&05 & 0&00 \\
11 & BETAC4\_O & 5&36 & 0&324 & 16&56 & 0&00 \\
12 & BETAC4\_R & 0&788 & 0&344 & 2&29 & 0&02 \\
\end{tabular}
\end{table}

\begin{table}[p]
\centering
\begin{tabular}{rlr@{.}lr@{.}lr@{.}lr@{.}l}
         &                       &   \multicolumn{2}{l}{}    & \multicolumn{2}{l}{\textbf{Robust}}  &     \multicolumn{4}{l}{}   \\
\textbf{Parameter} &                       &   \multicolumn{2}{l}{\textbf{Coeff.}}      & \multicolumn{2}{l}{\textbf{Asympt.}}  &     \multicolumn{4}{l}{}   \\
\textbf{number} &  \textbf{Description}                     &   \multicolumn{2}{l}{\textbf{estimate}}      & \multicolumn{2}{l}{\textbf{std. error}}  &   \multicolumn{2}{l}{\textbf{$t$-stat}}  &   \multicolumn{2}{l}{\textbf{$p$-value}}   \\
\hline
13 & BETAG1\_O & 0&502 & 0&268 & 1&87 & 0&06 \\
14 & BETAG1\_R & -0&487 & 0&358 & -1&36 & 0&17 \\
15 & BETAG2\_J & -0&619 & 0&210 & -2&95 & 0&00 \\
16 & BETAG2\_R & 0&0186 & 0&341 & 0&05 & 0&96 \\
17 & BETAG3\_J & -0&640 & 0&210 & -3&04 & 0&00 \\
18 & BETAG3\_O & -0&0184 & 0&264 & -0&07 & 0&94 \\
19 & BETAG4\_J & -0&685 & 0&210 & -3&26 & 0&00 \\
20 & BETAG4\_O & -0&0259 & 0&267 & -0&10 & 0&92 \\
21 & BETAG4\_R & 0&0429 & 0&347 & 0&12 & 0&90 \\
22 & BETAH1\_O & 0&823 & 0&270 & 3&05 & 0&00 \\
23 & BETAH1\_R & -7&45 & 0&419 & -17&78 & 0&00 \\
24 & BETAH2\_J & -0&0549 & 0&193 & -0&28 & 0&78 \\
25 & BETAH2\_R & -9&13 & 0&409 & -22&34 & 0&00 \\
26 & BETAH3\_J & 2&96 & 0&235 & 12&61 & 0&00 \\
27 & BETAH3\_O & 5&15 & 0&293 & 17&59 & 0&00 \\
28 & BETAH4\_J & 2&96 & 0&235 & 12&61 & 0&00 \\
29 & BETAH4\_O & 5&15 & 0&293 & 17&59 & 0&00 \\
30 & BETAH4\_R & 1&63e-05 & 0&341 & 0&00 & 1&00 \\
31 & BETAP1\_O & -5&36 & 0&281 & -19&10 & 0&00 \\
32 & BETAZ1\_R & -17&2 & 0&607 & -28&36 & 0&00 \\
33 & BETAZ2\_J & 2&24 & 0&202 & 11&08 & 0&00 \\
34 & BETAZ2\_R & -7&89 & 0&451 & -17&51 & 0&00 \\
35 & BETAZ3\_J & 4&72 & 0&259 & 18&24 & 0&00 \\
36 & BETAZ3\_O & 3&93 & 0&304 & 12&90 & 0&00 \\
37 & BETAZ4\_J & 4&61 & 0&262 & 17&62 & 0&00 \\
38 & BETAZ4\_O & 3&82 & 0&311 & 12&28 & 0&00 \\
39 & BETAZ4\_R & -0&105 & 0&383 & -0&27 & 0&78 \\
40 & BETAL1\_O & -7&23 & 0&399 & -18&13 & 0&00 \\
41 & BETAL1\_R & -14&1 & 0&581 & -24&26 & 0&00 \\
42 & BETAL2\_J & 4&12 & 0&280 & 14&73 & 0&00 \\
43 & BETAL2\_R & -3&73 & 0&339 & -10&99 & 0&00 \\
44 & BETAL3\_J & 4&68 & 0&288 & 16&28 & 0&00 \\
45 & BETAL3\_O & 1&92 & 0&246 & 7&78 & 0&00 \\
46 & BETAL4\_J & 4&83 & 0&313 & 15&44 & 0&00 \\
47 & BETAL4\_O & 3&24 & 0&285 & 11&37 & 0&00 \\
48 & BETAL4\_R & 3&36 & 0&342 & 9&81 & 0&00 \\
49 & BETAT1\_O & -3&57 & 0&255 & -13&98 & 0&00 \\
50 & BETAT1\_R & -13&2 & 0&531 & -24&91 & 0&00 \\
51 & BETAT2\_J & 0&459 & 0&167 & 2&75 & 0&01 \\
52 & BETAT2\_R & -6&28 & 0&414 & -15&17 & 0&00 \\
53 & BETAT3\_J & 1&72 & 0&253 & 6&81 & 0&00 \\
54 & BETAT3\_O & 2&98 & 0&290 & 10&28 & 0&00 \\
55 & BETAT4\_J & 4&55 & 0&387 & 11&74 & 0&00 \\
56 & BETAT4\_O & 6&50 & 0&436 & 14&90 & 0&00 \\
57 & BETAT4\_R & 2&52 & 0&528 & 4&77 & 0&00 \\
\hline
\end{tabular}
\end{table}

\clearpage
\subsection{Final Logit model}
\label{app_second_logit}

\begin{table}[h]
\centering
\begin{tabular}{rlr@{.}lr@{.}lr@{.}lr@{.}l}
         &                       &   \multicolumn{2}{l}{}    & \multicolumn{2}{l}{\textbf{Robus}t}  &     \multicolumn{4}{l}{}   \\
\textbf{Parameter} &                       &   \multicolumn{2}{l}{\textbf{Coeff.}}      & \multicolumn{2}{l}{\textbf{Asympt}.}  &     \multicolumn{4}{l}{}   \\
\textbf{number} &  \textbf{Description}                     &   \multicolumn{2}{l}{\textbf{estimate}}      & \multicolumn{2}{l}{\textbf{std. error}}  &   \multicolumn{2}{l}{\textbf{$t$-stat}}  &   \multicolumn{2}{l}{\textbf{$p$-value}}   \\

\hline
1 & ASC2 & -9&42 & 0&502 & -18&76 & 0&00 \\
2 & ASC3 & -1&97 & 0&317 & -6&23 & 0&00 \\
3 & ASC4 & 24&2 & 0&775 & 31&18 & 0&00 \\
4 & BETAC1\_O & -6&92 & 0&379 & -18&26 & 0&00 \\
5 & BETAC1\_R & -23&5 & 1&28 & -18&39 & 0&00 \\
6 & BETAC2\_J & 3&26 & 0&242 & 13&46 & 0&00 \\
7 & BETAC2\_R & -8&48 & 0&421 & -20&17 & 0&00 \\
8 & BETAC3\_J & 5&35 & 0&325 & 16&45 & 0&00 \\
9 & BETAC3\_O & 4&99 & 0&309 & 16&14 & 0&00 \\
10 & BETAC4\_J & 5&34 & 0&332 & 16&08 & 0&00 \\
11 & BETAC4\_O & 5&35 & 0&325 & 16&45 & 0&00 \\
12 & BETAC4\_R & 0&775 & 0&342 & 2&26 & 0&02 \\
13 & BETAG2\_J & -0&383 & 0&147 & -2&60 & 0&01 \\
14 & BETAG3\_J & -0&394 & 0&147 & -2&67 & 0&01 \\
15 & BETAG4\_J & -0&437 & 0&147 & -2&98 & 0&00 \\
16 & BETAH1\_O & 0&779 & 0&225 & 3&47 & 0&00 \\
17 & BETAH1\_R & -7&36 & 0&369 & -19&95 & 0&00 \\
18 & BETAH2\_R & -8&96 & 0&359 & -24&94 & 0&00 \\
19 & BETAH3\_J & 2&97 & 0&188 & 15&81 & 0&00 \\
20 & BETAH3\_O & 5&11 & 0&260 & 19&65 & 0&00 \\
21 & BETAH4\_J & 2&97 & 0&188 & 15&81 & 0&00 \\
22 & BETAH4\_O & 5&11 & 0&260 & 19&65 & 0&00 \\
23 & BETAZ1\_O & -5&35 & 0&280 & -19&13 & 0&00 \\
24 & BETAZ1\_R & -16&9 & 0&573 & -29&44 & 0&00 \\
25 & BETAZ2\_J & 2&21 & 0&198 & 11&17 & 0&00 \\
26 & BETAZ2\_R & -7&69 & 0&405 & -19&00 & 0&00 \\
27 & BETAZ3\_J & 4&65 & 0&241 & 19&28 & 0&00 \\
28 & BETAP3\_O & 3&88 & 0&273 & 14&18 & 0&00 \\
29 & BETAP4\_J & 4&59 & 0&243 & 18&89 & 0&00 \\
30 & BETAP4\_O & 3&84 & 0&277 & 13&89 & 0&00 \\
31 & BETAL1\_O & -7&20 & 0&397 & -18&15 & 0&00 \\
32 & BETAL1\_R & -13&9 & 0&572 & -24&30 & 0&00 \\
33 & BETAL2\_J & 4&08 & 0&275 & 14&87 & 0&00 \\
34 & BETAL2\_R & -3&66 & 0&333 & -11&00 & 0&00 \\
35 & BETAL3\_J & 4&64 & 0&283 & 16&40 & 0&00 \\
36 & BETAL3\_O & 1&91 & 0&244 & 7&84 & 0&00 \\
37 & BETAL4\_J & 4&78 & 0&307 & 15&57 & 0&00 \\
38 & BETAL4\_O & 3&24 & 0&284 & 11&42 & 0&00 \\
39 & BETAL4\_R & 3&30 & 0&339 & 9&72 & 0&00 \\
\end{tabular}
\end{table}

\begin{table}[h!]
\centering
\begin{tabular}{rlr@{.}lr@{.}lr@{.}lr@{.}l}
         &                       &   \multicolumn{2}{l}{}    & \multicolumn{2}{l}{\textbf{Robus}t}  &     \multicolumn{4}{l}{}   \\
\textbf{Parameter} &                       &   \multicolumn{2}{l}{\textbf{Coeff.}}      & \multicolumn{2}{l}{\textbf{Asympt}.}  &     \multicolumn{4}{l}{}   \\
\textbf{number} &  \textbf{Description}                     &   \multicolumn{2}{l}{\textbf{estimate}}      & \multicolumn{2}{l}{\textbf{std. error}}  &   \multicolumn{2}{l}{\textbf{$t$-stat}}  &   \multicolumn{2}{l}{\textbf{$p$-value}}   \\

\hline
40 & BETAT1\_O & -3&57 & 0&256 & -13&97 & 0&00 \\
41 & BETAT1\_R & -13&0 & 0&530 & -24&50 & 0&00 \\
42 & BETAT2\_J & 0&451 & 0&167 & 2&70 & 0&01 \\
43 & BETAT2\_R & -6&16 & 0&410 & -15&00 & 0&00 \\
44 & BETAT3\_J & 1&69 & 0&246 & 6&87 & 0&00 \\
45 & BETAT3\_O & 2&97 & 0&286 & 10&35 & 0&00 \\
46 & BETAT4\_J & 4&50 & 0&379 & 11&86 & 0&00 \\
47 & BETAT4\_O & 6&47 & 0&430 & 15&04 & 0&00 \\
48 & BETAT4\_R & 2&53 & 0&520 & 4&86 & 0&00 \\
\hline
\end{tabular}
\end{table}

\clearpage
\subsection{Parameters of the retained neural network}
\label{app_neural_net}

\subsubsection{Weights of the links connecting the input nodes to the hidden nodes}

\begin{table}[h!]
\centering
  \begin{tabular}{c|rrrrrrr}

\textbf{From \textbackslash To}  & \multicolumn{1}{c}{H1} & \multicolumn{1}{c}{H2} & \multicolumn{1}{c}{H3} & \multicolumn{1}{c}{H4} & \multicolumn{1}{c}{H5} & \multicolumn{1}{c}{H6} & \multicolumn{1}{c}{H7}\\
\hline     
I1  & -35.867 & 23.658  &  2.391 & 11.670  & 27.745  &  -13.459  &   19.404\\
I2  & -35.996 & 23.658  &  2.454 & 11.653  & 27.338  &  -13.396   &  19.454\\
I3  &  8.614 & -13.820 &  -2.370 & -55.420  & -9.719    &  1.004  & -165.126\\
I4   &  8.632 & -13.853  & -2.440 & -55.902  & -9.725   &   1.024  & -164.843\\
I5 & -110.255 & 25.017  &  1.840 & 102.037 & -129.940  & -725.158  &  173.844\\
I6 & -53.380 & 25.829  & -0.154 & 24.625   &  78.973  & -725.149  &   49.021\\
I7 &  15.503 & -11.158  & -4.433 & -65.232   &  10.124  &   -0.027  &  -51.453\\
I8 &  15.570 & -11.256  & -4.329 & -66.214   &  10.154   &  -0.230  &  -51.364\\
I9 & -11.635  & -0.391  & -1.761  & -3.249  &   28.207  &    0.978  &   16.604\\
I10 &  -0.120 & -0.114  & -1.258  & 1.076   &   0.067   &   0.559   &  -1.045\\
I11 & -0.114  & -0.165  & -1.207  & 0.320   &   0.079  &    0.536   &  -0.966\\
I12 &  -0.070 & -0.077  & -1.298  & 0.124   &   0.093   &   0.563   &  -1.098\\
I13 & -861.710  &  8.578  &  4.513 & 159.844  &  331.590  &  -15.664  & -945.908\\
I14 & -87.364 & -12.510  &  0.001  & 0.600  &  -17.525   &   2.097  &   17.449\\
I15 &  13.419 &  0.001  & -2.331 & -63.573  &  130.627  &  -15.933  & -107.060\\
I16 &  13.570 & -0.032  & -2.305 & -76.966  &  130.403  &  -15.879  & -106.733\\
I17 & -34.691 & -3.290   & 3.121 & 113.275  &   -0.647  &   -4.096  &  -53.385\\
I18  &  0.002  &  23.614  &  1.756  & 2.115   &  11.846  &   10.384  &   18.817\\
I19 &  67.795  &  83.160  &  0.767 & -15.354   &  63.344  &   -5.547  &    5.041\\
I20 &  80.525  &  83.227  & -2.144 & -42.375  &  382.170   &  -7.364  &  -51.399\\
I21 & -62.784  &  -9.082  &  3.572 & 146.629  & -186.594  & -724.555  &   -3.171\\
I22 &  -3.226  &  20.698  & -0.499 &  0.923  &   21.966   &   0.606   &   7.002\\
I23 & 13.412  &  32.972  & -1.244 & -33.775   &  22.012   &   0.167   &   0.717\\
I24  & 47.062  &  77.420  & -1.227 & -57.389  &  116.664  &   -8.501  &  -24.739\\
  \end{tabular}
\end{table}

\subsubsection{Activation threshold of the hidden nodes}

\begin{table}[h!]
\centering
  \begin{tabular}{c|ccccccc}
    & H1 & H2 & H3 & H4 & H5 & H6 & H7\\
    \hline
$\theta_i$   &    0.157 &  0.670  & -1.189 &  -0.628  &  -1.685 &   0.238 &   0.370
  \end{tabular}
\end{table}

\subsubsection{Weights of the links connecting the hidden nodes to the output nodes}

\begin{table}[h!]
\centering
  \begin{tabular}{c|rrrr}
  \textbf{From \textbackslash To}  & \multicolumn{1}{c}{O1} & \multicolumn{1}{c}{O2} & \multicolumn{1}{c}{O3} & \multicolumn{1}{c}{O4}\\
  \hline
H1 & -0.015 & -0.019 & -0.893 & 0.928\\
H2 &  0.411 & -0.845 &  0.246 & 0.187\\
H3 &  0.982 & -0.916 & -0.032 & -0.033\\
H4 & -0.081 &  0.897 & -0.832 & 0.017\\
H5 & -0.539 &  0.748 & -0.111 & -0.097\\
H6 &  0.169 &  0.087 & -1.028 & 0.771\\
H7 & -0.456 &  0.741 & -0.189 & -0.095\\
  \end{tabular}
\end{table}

\end{document}